\documentclass[
    ,final            
  ]
  {aipproc}

\layoutstyle{6x9}

\begin{document}

\title{Mass Distributions Beyond TDHF}


\author{J.~M.~A.~Broomfield}{
  address={Department of Physics, University of Surrey, Guildford, Surrey, GU2 7XH. UK}
}

\author{P.~D.~Stevenson}{
  address={Department of Physics, University of Surrey, Guildford, Surrey, GU2 7XH. UK}
}


\begin{abstract}
The mass distributions for giant dipole resonances in $^{32}$S and $^{132}$Sn decaying through particle emission and for 
deep-inelastic collisions between $^{16}$O nuclei have been investigated by implementing the  
Balian-V\'{e}n\'{e}roni variational technique based upon a three-dimensional time-dependent Hartree-Fock code with realistic Skyrme 
interactions. The mass distributions obtained have been shown to be significantly larger than the standard
TDHF results.
\end{abstract}

\maketitle


\section{Introduction}

The standard time-dependent mean-field methods used in nuclear physics (namely the time-dependent Hartree-Fock (TDHF) 
approach and its derivatives \cite[]{cite:lacroix,cite:ayik}) have been successfully used to determine the 
expectation values for single-particle observables, such as fragment 
mass, in nuclear reactions and decays but severely underestimate the fluctuations in these values \cite{cite:reinhard}. This was 
first observed by Davies et al.~\cite{cite:davies} in 1978 who performed TDHF calculations to investigate the 
full-width-half-maximum of the mass of the projectile-like
fragment for the heavy-ion reactions $^{84}$Kr+$^{208}$Pb ($E_{lab}$=$494$\,MeV) and $^{84}$Kr+$^{209}$Be ($E_{lab}$=$600$\,MeV) 
and found that they underestimated the experimental results by about an order of magnitude. It is now well known that the 
TDHF method underestimates the fluctuation of any single-particle operator, or
in fact the expectation value of any many-body
operator, except the energy \cite{cite:reinhard}. This is a
consequence of the central assumption of the methods; the
approximation of the full many-body state of the system as a Slater
determinant, evolving in a mean-field, neglecting explicit two-body correlations.

This problem was studied by Balian and V\'{e}n\'{e}roni
\cite{cite:bv1,cite:bv2,cite:bv3} in the 1980's, who produced a
general variational theory optimised to the determination of the
expectation values and fluctuations for arbitrary single-particle operators 
whilst the state of the system is given by a single Slater determinant. 
They found that, for a system described, at the time $t_0$, by the
one-body density matrix, 
$\rho\left(t_0\right)$, the expectation value for the single-particle observable $\hat{Q}$, at the later time $t_1$, is given 
by $\left.\langle Q\rangle\right|_{t_1}=\mbox{Tr}\left[\hat{Q}\rho\left(t_1\right) \right]$,
in keeping with the usual TDHF approach, whilst the distribution, or standard deviation, $\Delta Q$, is given by
\begin{equation}
\left. \left( \Delta Q_{BV} \right)^2 \right|_{t_1}
  = \lim\limits_{\varepsilon \to 0} \frac{1}{2\varepsilon^2}
     \mbox{Tr}\left[ \rho\left(t_0\right) - \sigma\left(t_0,\varepsilon\right) \right],
\label{eqn:bv}
\end{equation}
where $\sigma\left(t,\varepsilon\right)$ is a one-body density matrix related to $\rho\left(t\right)$ through the boundary
condition
\begin{equation}
\sigma\left(t_1,\varepsilon\right)
  = \exp\left(\mbox{i}\varepsilon \hat{Q}\right)\rho\left(t_1\right) \exp\left(-\mbox{i}\varepsilon\hat{Q}\right), \label{eqn:sigma}
\end{equation}
The time evolution of $\rho\left(t\right)$ and $\sigma\left(t,\varepsilon\right)$ is given by the
usual TDHF equation. This latter result differs from the usual TDHF result
\begin{equation}
\left. \left( \Delta Q_{TDHF} \right)^2 \right|_{t_1}
   = \mbox{Tr} \left[ \hat{Q} \rho\left(t_1\right) \hat{Q} \left( 1 - \rho\left(t_1\right) \right) \right], \label{eqn:tdhf}
\end{equation}
in that it depends explicitly on the initial time, $t_0$, with the final time, $t_1$, entering 
through the boundary condition (\ref{eqn:sigma}). The boundary condition (\ref{eqn:sigma}) also contains
the operator $\hat{Q}$ such that this approach is specifically tuned
to the determination of $\Delta Q$.  It has also been shown by Dasso \cite{cite:dasso} that, for
simple operators such as mass (or, equivalently, charge), 
the single-particle nature of TDHF leads to an unphysical upper limit on the mass distributions that can be obtained from (\ref{eqn:tdhf}). 
It can be shown that
\begin{equation}
\left.\left(\Delta N_{TDHF}^{2}\right)_{MAX}\right|_{t}
  = \left.\langle N\left(R_c\right)\rangle\right|_{t}\left(1-\frac{\left.\langle N\left(R_c\right)\rangle\right|_{t}}{A}\right),
\label{eqn:max}
\end{equation}
where $A$ is the total number of nucleons in the system. 

Solving (\ref{eqn:bv}) requires that a
Hartree-Fock calculation be performed to determine the initial state,
$\rho\left(t_0\right)$ followed by a suitable instantaneous excitation of the system 
and a TDHF calculation from $t_0$ to $t_1$ to determine
$\rho\left(t_1\right)$. The transformation (\ref{eqn:sigma})
gives $\sigma\left(t_1,\varepsilon\right)$ and a
second TDHF calculation must then be performed, backwards from $t_1$ to $t_0$, to obtain
$\sigma\left(t_0,\varepsilon\right)$. These latter steps must be
repeated for a range of values of $\varepsilon$ to allow $\Delta
Q_{BV}$ to be determined from the extrapolation of $\varepsilon$ to $0$.

The large number of calculations required to evaluate (\ref{eqn:bv}),
and their complexity, means that only a handful
of calculations were performed using this method
and those calculations used simplified
interactions and symmetries (either spherical
\cite{cite:troudet}, or axial \cite{cite:marston,cite:bonche}) to
make the problems solvable. However, advances in computing power
mean that this approach can now be implemented using fully
three-dimensional TDHF codes with full Skyrme interactions
\cite{cite:maruhn,cite:umar,cite:simenel,cite:nakatsukasa,cite:broomfield}.

Written in terms of the occupied single-particle states (\ref{eqn:bv}) becomes \cite{cite:marston}
\begin{equation}
\left. \left( \Delta N_{BV} \right)^2 \right|_{t_1}
  = A - \lim\limits_{\varepsilon \to 0} \frac{f\left(\varepsilon\right)}{\varepsilon^2}, \label{eqn:sp-bv} \\
\end{equation}
where
\begin{equation}
f\left(\varepsilon\right)
  = \sum\limits_{m,n<\epsilon_F} \int \mbox{d}\bar{r}
    \left| \left< \psi_m\left(\bar{r},t_0,\varepsilon\right) \right.\left| \phi_n\left(\bar{r},t_0\right) \right> \right|^2,
\end{equation}
and $A$ is the number of nucleons in the system. 
The wavefunctions $\left|\psi_m\left(\bar{r},t,\varepsilon\right)\right>$ are the wavefunctions
from the backwards TDHF calculations and are related to the wavefunctions from the forwards calculations,
$\left|\phi_n\left(\bar{r},t\right)\right>$, through the boundary condition
\begin{equation}
\psi\left(\bar{r},t_1,\varepsilon\right)
  = \exp\left(\mbox{i}\varepsilon \hat{Q} \right) \phi\left(\bar{r},t_1\right).
\label{eqn:sp-psi}
\end{equation}
In these calculations, as in Bonche and Flocard's earlier work \cite{cite:bonche}, the wavefunctions $\left|\phi_n\left(\bar{r},t_0\right)\right>$ 
were obtained by evolving the TDHF equations forwards and then backwards
without the transformation (\ref{eqn:sp-psi}) being applied. This approach ensures that all the single-particle wavefunctions 
used in evaluating (\ref{eqn:sp-bv}) result from the same number of computations and has been found to significantly reduce systematic 
numerical errors \cite{cite:broomfield} allowing the extrapolation required in (\ref{eqn:sp-bv}) to be extended
to significantly smaller values of $\varepsilon$. 

We consider the mass distribution in a bounded region of space around
a nucleus and calculate the mass (number of nucleons) in the nucleus according to 
\begin{equation}
\left. \langle N\left(R_c\right) \rangle \right|_{t}
  = \sum\limits_{m<\epsilon_F} \int \mbox{d}\bar{r}\left|\phi_m\left(\bar{r},t\right)\right|^2 
  \theta\left(R_c-\left|\left(\bar{r}-\bar{r}_{CM}\right)\right|\right),
\label{eqn-n}
\end{equation}
where $R_c$ is the cut-off radius used to define the bounded region of
space and $\bar{r}_{CM}$ is the centre-of-mass location of the nucleus.


\section{GDRs in $\mathbf{^{32}}$S and $\mathbf{^{132}}$Sn}

We consider first a giant dipole resonance (GDR) in the deformed nucleus $^{32}$S decaying
through particle emission. The
calculations were all performed in a cubic spatial box of size
$32\times 32\times 32$\,fm discretised in steps of $1$\,fm.  The
static Hartree-Fock calculation was carried out using the SLy6  
parametrisation \cite{cite:chabanat} of the Skyrme interaction and produced 
a ground state with prolate deformation ($\beta_2=0.11$).  A dipole excitation was induced by
acting on each wavefunction at $t=0$ with a boost
\begin{equation}
B_{D}\left(x,y,z\right)
  = \exp\left( \mbox{i} F C \left( A_x x + A_y y + A_z z\right) \right), \label{eqn:gdr-boost}
\end{equation}
where
\begin{equation}
C = \sqrt{\frac{5}{4\pi}} \frac{1}{1+\exp\left(\sqrt{x^2+y^2+z^2}\right)}, \nonumber
\end{equation}
is a spatial cut-off, and where, for protons, $F=1/Z$, and for neutrons, $F=-1/(A-Z)$, where $A$ is the nucleus' atomic mass 
and $Z$ is its charge. The $A_x$, $A_y$ and $A_z$ parameters determine
the strength of the applied boost and were set to $112.5$\,fm$^{-1}$.

In the present calculation, the system was evolved forward in time until $t_1=250$ fm/c.  
Dirichlet boundary conditions were used at the edge of the spatial box, where the wavefunctions disappear. 
These lead to spurious reflections for sufficiently large $t_1$ so the value chosen for $t_1$ had to be kept sufficiently small,
whilst ensuring that $\left.\langle N\left(R_c\right)\rangle\right|_t$ had stabilised
following the prompt de-excitation of the resonance by particle emission.

\begin{figure}
\rotatebox{-90}{\includegraphics[width=.35\textwidth]{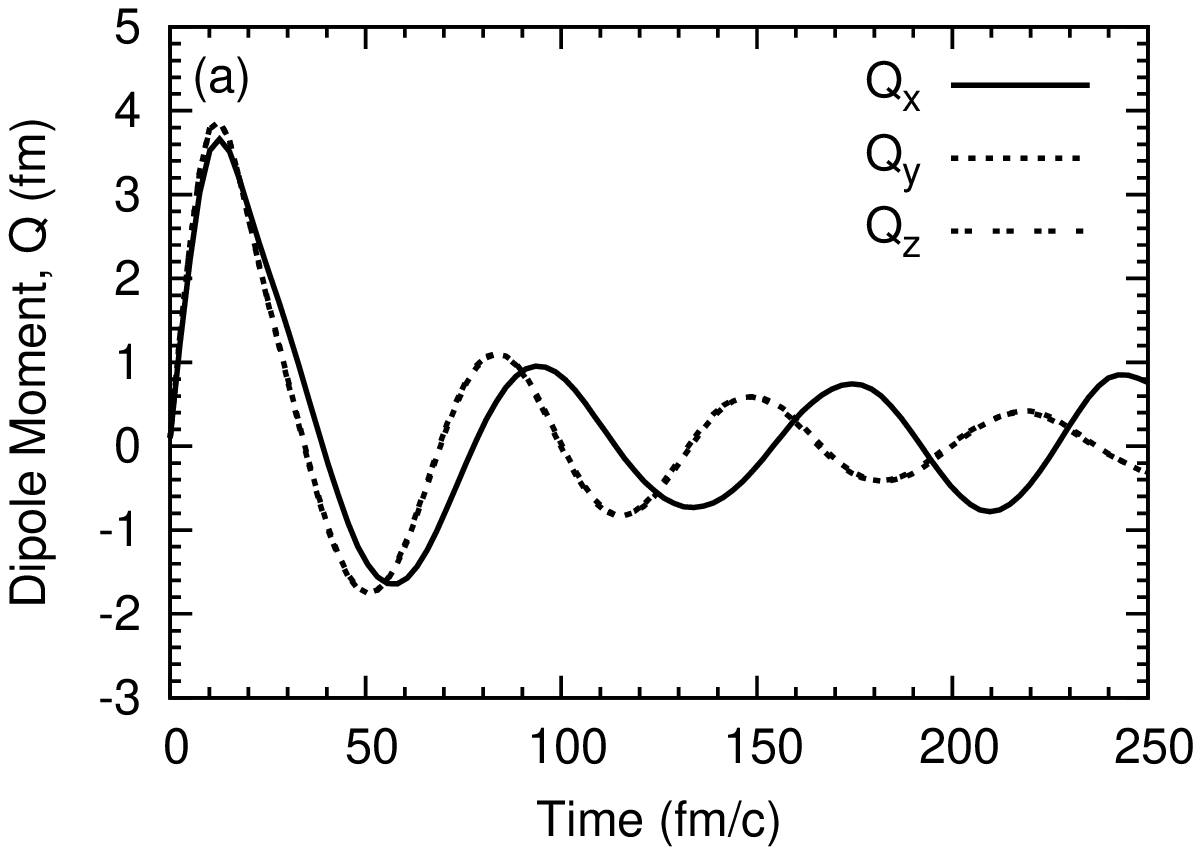}}
\rotatebox{-90}{\includegraphics[width=.35\textwidth]{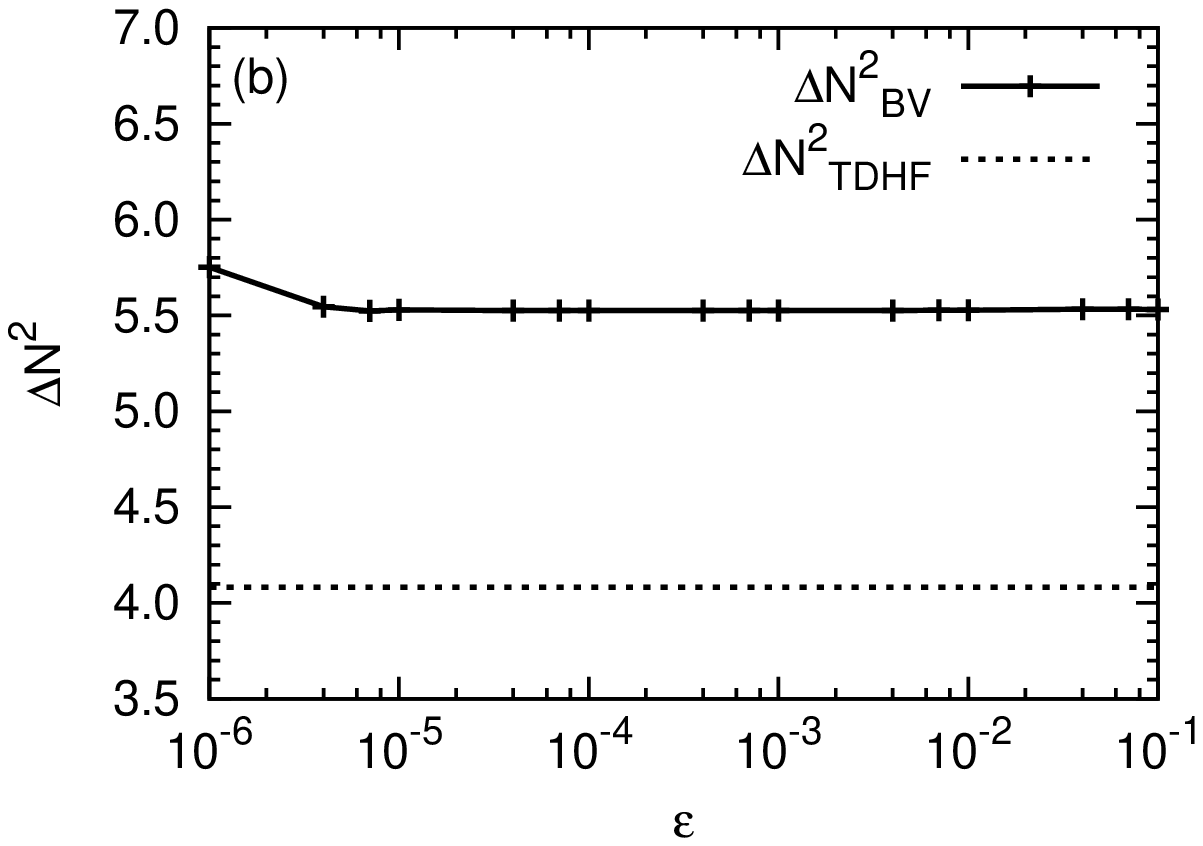}}
\caption
  {(a) The dipole moments $Q_i$ ($i$=$x$, $y$, $z$) as a function of time, for a GDR in $^{32}$S. The differences between
  $Q_x$ and $Q_y$ and $Q_z$ are consistent with a prolate deformed ground state where $x$ is the
  long axis. (b) $\Delta N^2_{BV}$ as a function of $\varepsilon$. $\Delta N_{BV}^{2}$ is given by extraplation to 
  $\varepsilon\to 0$. The $x$-axis has a logarithmic scale to emphasise that the values obtained are independent 
  of $\varepsilon$ across several orders of magnitude. 
  The standard TDHF result (calculated at $t_1$ and independent of $\varepsilon$) is shown for reference.}
\label{fig:s32}
\end{figure}

The dipole moments, $Q_x$, $Q_y$ and $Q_z$, as a function of time, are
shown in figure \ref{fig:s32}(a), in accordance with \cite{cite:simenel}
\begin{equation}
Q_i = \frac{\left(A-Z\right)Z}{A} \left( \langle x_i^P \rangle - \langle x_i^N \rangle \right),
\label{eqn:dip}
\end{equation}
where $i=1$, $2$, $3$ denotes $x$, $y$ and $z$ and $\langle x_i^P\rangle$ and $\langle x_i^N\rangle$ are the expectation values for
position calculated using the proton and neutron single-particle states respectively. 
The prolate deformation of the $^{32}$S nucleus (with $x$ the long-axis)
results in $Q_y$ and $Q_z$ being identical whilst $Q_x$ differs. The periodicity of $Q_x$, $Q_y$ and
$Q_z$ provides an estimate of the excitation energy for the
oscillations along each of the three primary axes.  For $Q_x$, a
period of $\approx 71$\,fm/c is found, giving an excitation energy $E_x \approx 17.5$\,MeV and, for $Q_y$ and $Q_z$, a
period of $\approx 68$\,fm/c gives an excitation energy $E_y\approx E_z\approx\,18.3$\,MeV.

Following the decay of the GDR in $^{32}$S, it was found that $\langle
N \rangle=26.65$ and  $\Delta N_{TDHF}^2=4.08$ (using $R_c=8$\,fm)
representing the emission of $\approx 5$ nucleons. $R_c$ was chosen so
that the bounded region fully enclosed the nucleus but omitted, as
much as possible, the 
radiated components of the wavefunctions. From (\ref{eqn:max}) we
obtain $\left(\Delta N_{TDHF}^{2}
\right)_{MAX}=4.46$. Following evolution to $t_1$, the transformation
(\ref{eqn:sp-psi}) was applied and the system evolved back to $t_0$ for
$\varepsilon$ values down to $10^{-6}$. After each calculation the
fluctuation, $\Delta N_{BV}^2\left(\varepsilon\right)$, was
calculated using (\ref{eqn:sp-bv}) and the results plotted 
and extrapolated back to $\varepsilon=0$. The
results are shown in figure \ref{fig:s32}(b), giving 
$\Delta N_{BV}^2=5.52$ which represents a $16$\% increase in
$\Delta N$ and exceeds the TDHF upper limit, $\left( \Delta N_{TDHF}^{2}
\right)_{MAX}$. Calculations were also performed for $R_c=8.5$\,fm and
$R_c=9$\,fm and showed only small changes in the results consistent
with the bounded region enclosing increasing amounts of the
wavefunctions tails.
%

The calculations were repeated for the doubly magic nucleus
$^{132}$Sn using the same spatial box and interaction as in the
$^{32}$S calculation. 
An initial dipole
boost (\ref{eqn:gdr-boost}) with $A_x=A_y=A_z=600$\,fm$^{-1}$ induced
the resonance, which was evolved from $t_0=0$\,fm/c to $t_1=250$\,fm/c 
as before. The dipole moments show a periodicity of $\approx 88$\,fm/c 
corresponding to a resonance energy of $\approx 14.1$\,MeV which is close to
the experimental value of $16.1\left(7\right)$\,MeV \cite{cite:adrich}.

The standard THDF calculation gave, at the time $t_1$, $\langle
N\rangle=121.02$ and $\Delta N_{TDHF}^{2}=8.46$ representing the 
emission of $11$ nucleons. From (\ref{eqn:max}) we obtain $\left(
  \Delta N_{TDHF}^{2} \right)_{MAX}=10.07$. 
Again, the Balian-V\'en\'eroni transformation was applied for different
values of $\varepsilon$, the extrapolation to $\varepsilon=0$ yielding 
$\Delta N_{BV}^{2}=11.29$, representing a $14$\% increase in $\Delta N$ compared with the standard TDHF result.

\section{$\mathbf{^{16}}$O+$\mathbf{^{16}}$O Collisions ($\mathbf{E_{CM}=128}$\,MeV)}

The main application for this method will be heavy-ion collisions as
mentioned in the introduction. As a test case using light nuclei,
collisions between $^{16}$O nuclei at $E_{CM}=128$\,MeV have been
studied. 

The dynamic calculations were carried out 
in a rectangular space of size $50\times 32\times 50$\,fm except the head-on calculation where, taking advantage of 
the symmmetry a spatial box of size $50\times 32\times 32$\,fm was used. 
The nuclei were initially positioned at $\left(\pm 10,0,\pm b/2\right)$ 
where $b$ is the impact parameter  and the nuclei collide in the $x$-$z$ plane. 
\begin{figure}
\rotatebox{0}{\includegraphics[width=1.00\textwidth]{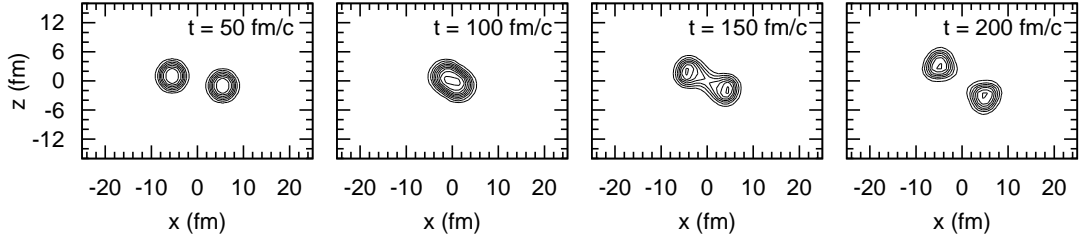}}
\caption
  {Density contour plots showing a deep inelastic collision ($E_{CM}=128$\,MeV) between two $^{16}$O nuclei 
  for an impact parameter $b=2$\,fm ($l\approx 14\mbox{\,}\hbar$).}
\label{fig:o16}
\end{figure}

Calculations were performed for impact parameters $b$ = $0$, $1$, $2$ and $4$\,fm. 
Contour plots showing the densities during a collision for $b=2$\,fm are shown in figure \ref{fig:o16}. The calculations were run
for $2000$\,timesteps ($400$\,fm/c) until the scattered nuclei had clearly separated but before they reached the edge of
the box. The exception was the calculation for $b=4$\,fm where the nuclei fused with the compound nucleus decaying through 
particle emission and where the runtime was extended to $2000$\,fm/c
to allow the excited compound nucleus time to decay and check that it did not undergo a delayed fission. 
\begin{table}[b]
\begin{tabular}{ccccccc}
\hline
\tablehead{1}{c}{c}{b (fm)}
  & \tablehead{1}{c}{c}{$\mathbf{l}$ ($\mathbf{\hbar}$)}
  & \tablehead{1}{c}{c}{$\mathbf{\left.\langle N\left(R_c=8\mbox{\,fm}\right) \rangle \right|_{t_{1}=400\mbox{\,fm/c}}}$}
  & \tablehead{1}{c}{c}{$\Delta\mathbf{N_{TDHF}}$}
  & \tablehead{1}{c}{c}{$\mathbf{\left(\right.}\Delta\mathbf{\left.N_{TDHF}\right)_{MAX}}$}
  & \tablehead{1}{c}{c}{$\Delta\mathbf{N_{BV}}$}
  & \tablehead{1}{c}{c}{Change} \\ \hline
0 &  0.00 & 14.96 & 1.37 & 2.82  & 2.84 & +107\% \\
1 &  7.02 & 14.90 & 1.38 & 2.82  & 2.87 & +108\% \\
2 & 14.05 & 14.83 & 1.44 & 2.82  & 3.17 & +120\% \\ 
4 & 28.10 & 25.10\tablenote{After $2000$\,fm/c. The $^{16}$O nuclei fused to form an excited 
  compound nucleus which decays by particle emission. The reduced value of  
  $\left(\Delta N_{TDHF}\right)_{MAX}$ for $b=4$\,fm is a consequence of the 
  dependence of (\ref{eqn:max}) on both the number of nucleons in the system, $A$, and 
  on the number of nucleons in the nucleus of interest, $\left.\langle N\left(R_c\right)\rangle\right|_{t}$.} 
  & 2.06 & 2.33  & -\tablenote{Not calculated due to the extended runtime required for the compound nucleus to decay and to ensure 
  that the compound nucleus did not fission.} 
  & - \\ \hline
\end{tabular}
\caption{Mass distributions calculated for $^{16}$O+$^{16}$O at $E_{CM}=128$\,MeV using the standard TDHF 
  approach and the Balian-V\'{e}n\'{e}roni variational approach for a range of impact parameters (the 
  equivalent angular momentums are also given). }
\label{tab:o16}
\end{table}

The mass distributions obtained from these calculations using the TDHF
and BV approaches are given in table \ref{tab:o16}, showing the large
increase (at least doubling in each case) of $\Delta N$ in the BV case
over TDHF.

\section{Conclusions}

The Balian-V\'{e}n\'{e}roni approach has been implemented using a three-dimensional TDHF code with the full Skyrme interaction and
calculations have been performed for GDR's in $^{32}$S and $^{132}$Sn and for collisions of $^{16}$O nuclei. 
The BV approach produces mass distributions which are quantitatively larger than those obtained using the usual TDHF approach.
We are continuing to apply this approach to heavy ion collisions with a view towards performing calculations for heavier
systems to allow a comparison with experimental data.


\begin{theacknowledgments}
The authors are pleased to acknowledge the assistance of, and discussions with,
Ph.~Chomaz, R.~Balian, J.~S.~Al-Khalili and E.~B.~Suckling. 
This work was supported by the UK Science and Technology Facilities Council (STFC). 
\end{theacknowledgments}



\bibliographystyle{aipproc}   





\end{document}